\documentclass[9pt]{article}
\usepackage{spconf,amsmath,graphicx}
\usepackage{url}
\usepackage{microtype}
\usepackage[english]{babel}
\usepackage{blindtext}
\usepackage [autostyle]{csquotes}
\usepackage[symbol]{footmisc}

\MakeOuterQuote{"}

\title{Relating Human Perception of Musicality to Prediction in a Predictive Coding Model}

\begin{document}


\name{%
\begin{tabular}{@{}c@{}}
Nikolas McNeal$^{1 *}$ \qquad 
\thanks{*\quad Equal contribution}
Jennifer Huang$^{2 *}$ \qquad 
Aniekan Umoren$^{3}$ \qquad
Shuqi Dai$^{2}$ \\
Roger Dannenberg$^{2}$ \qquad 
Richard Randall$^{2}$ \qquad 
Tai Sing Lee$^{2}$
\end{tabular}}




\address{$^{1}$ The Ohio State University \\ 
         $^{2}$ Carnegie Mellon University \\
         $^{3}$ Massachussetts Institute of Technology \\
         }

\maketitle



%
%
%
%
%
%
\begin{abstract}
We explore the use of a neural network inspired by predictive coding for modeling human music perception. This network was developed based on the computational neuroscience theory of recurrent interactions in the hierarchical visual cortex. When trained with video data using self-supervised learning, the model manifests behaviors consistent with human visual illusions. Here, we adapt this network to model the hierarchical auditory system and investigate whether it will make similar choices to humans regarding the musicality of a set of random pitch sequences. When the model is trained with a large corpus of instrumental classical music and popular melodies rendered as mel spectrograms, it exhibits greater prediction errors for random pitch sequences that are rated less musical by human subjects. We found that the prediction error depends on the amount of information regarding the subsequent note, the pitch interval, and the temporal context. Our findings suggest that predictability is correlated with human perception of musicality and that a predictive coding neural network trained on music can be used to characterize the features and motifs contributing to human perception of music.
\footnote{Our source code is available at \url{https://github.com/leelabcnbc/predictive-coding-music-prediction}}

\end{abstract}
\section{Introduction}
\label{sec:intro}

Over the past few decades, deep neural networks have made significant progress in becoming effective tools for solving various practical tasks, from image classification to speech-to-text translation. The representations of neuron units in such task-driven networks have been found to be remarkably similar to those of human and non-human primates \cite{yamins_using_2016}. One particular hierarchical recurrent neural network, PredNet, \cite{PredNet} is inspired by the predictive coding theory \cite{raoballard} for modeling hierarchical interaction in the visual cortex. This model learns an internal representation of the world through self-supervised learning. It generates a prediction of the incoming signals and uses the discrepancy of these predictions against the actual incoming signals as an error signal to drive the learning process. The network is modeled after the hierarchical inference process in the perceptual systems of the primate brain, with each level of representation modeling the bottom-up representation while simultaneously trying to align with the top-down expectation from above. 
Interestingly, this model has been shown to exhibit behaviors reminiscent of human perceptual visual illusions \cite{WatanabeEiji2018IMRb} as well as neurons' behaviors at different levels of the primate visual systems  \cite{4499}. In this paper, we explore whether this class of predictive coding-inspired neural network models can be used to model aspects of human music perception. A neural network model that can predict human perceptual experience can be used to characterize the complex dynamics and representations behind musicality perception.

In music, the relationship between musicality and prediction has long been recognized as an integral part of the music experience. A number of studies have tried to relate musicality to expectation with measures of predictability \cite{Huron_2013, uncertaintysurprisebrain, predocdingismir, palmer}. While these measures are intuitive and interpretable, they are not able to capture the entire complexity of expectation or prediction. Music is governed by organizational regularities. As such, repeated exposure to these regularities allows listeners to form expectations about what the next note in a melody might be, conditioned on what they have previously heard \cite{huron_2008}. 
Here, we propose that a predictive coding neural network might help model the complexity of experience for prediction, and we hypothesize that the model's predictability of pitch sequences is correlated with human perception of musicality. 


\begin{figure}[h]
    \begin{center}
    \includegraphics[width=1\columnwidth, height=140pt]{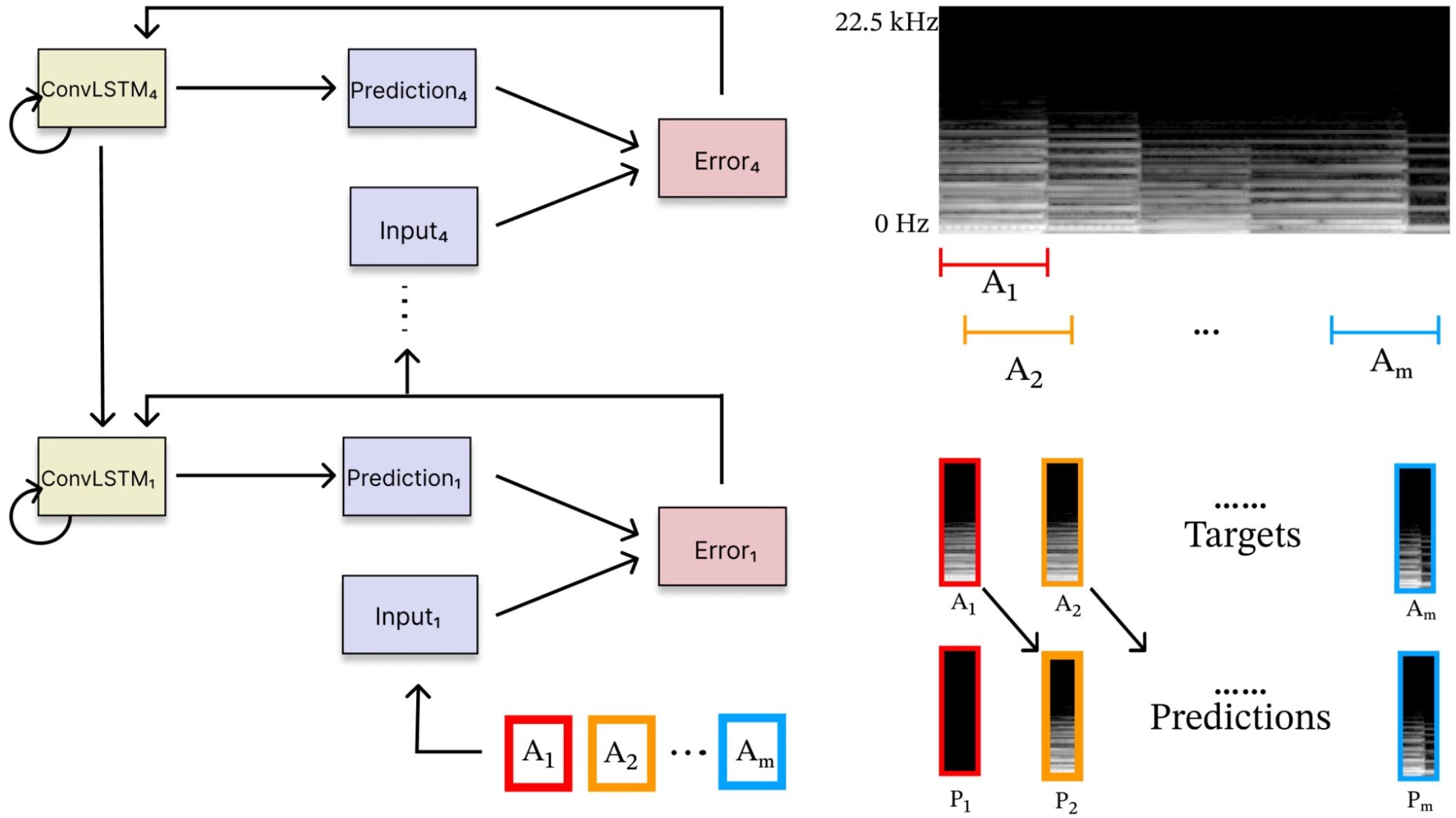}
    \centering
    \end{center}
    \caption{\label{fig:frog1} Left: Architecture of PredNet and the input sequences of mel spectrogram frames. Right: Mel spectogram of a musical sequence. Frame $A_{n}$ represents a 128 x 44 segment extracted to input to the PredNet at each time. Frame $P_{n+1}$ is the predicted frame given $A_{n}$. The prediction error between the prediction $P_{n+1}$ and the target $A_{n+1}$ is used to train the network.}
\end{figure}

\section{Approach}
\label{sec:approach}

Deep learning models have demonstrated their ability to capture statistical regularities and implicit rules of music \cite{shuqidannenberg, surprisenet, one, two} as well as complex features of speech and audio \cite{apa, cpc,wav2vec}. We use PredNet, a machine learning model from the video prediction literature, in developing our music prediction model. 
Significant advances in next-frame video prediction have been made since the introduction of PredNet. While these modern architectures would yield more accurate frame predictions, PredNet remains the simplest deep learning model of this class that mirrors the hierarchical organization of the sensory systems in the brain that are relatable to the human experience \cite{WatanabeEiji2018IMRb}. As such, the model allows us to explore hierarchical representations that are meaningful for making predictions.

We trained the network with the Medley-solos-DB dataset \cite{medleysolosdb}, which consists of 8,053 single instrument audio recordings of 2,972 ms length. This dataset is chosen for its simplicity, as it consists of monophonic melodies.
We also trained the model using the single-instrument tracks from the URMP dataset with clips of ten seconds length. We found that the two models did not yield a  significant difference in performance in comparing with human subjects' musicality judgment of our test sequences, so we will primarily focus discussion on the model trained with the Medley-solos-DB dataset.
We trained the network with acoustic music signals, converted in mel spectrogram format,  rather than symbolic data (such as MIDI) as is commonly done in music generation studies, in order to better simulate human experience. In addition, training with acoustic signals allows us to endow the network with greater flexibility to learn more complex and feature-rich music, and thus makes the model more generalizable to out-of-domain data.


Mel spectrograms are commonly used in speech recognition and music classification for use in convolutional neural networks \cite{costa}. The mel spectrogram is a spectral-temporal representation of sound, with frequency in the mel scale represented on the y-axis and time represented on the x-axis. The mel scale, a perceptual scale in which humans perceive pitches to be equally distanced from each other, is used to more closely simulate human perception of the audio. Each 3-second clip in the Medley-solos-DB dataset is converted into a mel spectrogram of dimensions \( \mathcal{C} \) x \( \mathcal{H} \) x \( \mathcal{W} \), where \( \mathcal{H} \)=128 (the number of frequency bands), and \( \mathcal{W} \)=257 (time units at 11.56 ms interval). The frequencies cover 0 Hz to 22,050 Hz. 

A 128 x 44 segment of the mel spectrogram is input to PredNet as a frame, meaning that the system processes a clip of 508 ms of signal at a time. This allows the network's convolutional units to learn spectro-temporal patterns in that local context. The network, through its convLSTM units \cite{convlstm}, learns how these patterns evolve over frames. The duration of the local temporal context is a hyperparameter. We experimented with a range of values for this parameter, from 0.25 seconds to 0.75 seconds, and found 0.5 seconds optimal in the context of the musicality experiment described below. The power for a given frequency at a given time in the mel spectogram is converted from a relative scale of -80 to 0 dB  to [0,255] pixel value. We compute our prediction errors using mean squared error (MSE) directly from the [0, 255] values.

The basic model we use is composed of a hierarchy of four convLSTM layers stacked on top of each other, though we also experimented with training PredNet with a different number of layers to assess the contribution of the higher layers for learning global temporal context and to ensure reproducibility of the results. At the bottom layer of PredNet, the network generates a predicted spectrogram frame, which is compared with the input frame. The error signal feeds forward to the next level as well as into the convLSTM that generates the prediction. This next level's convLSTM makes a prediction and compares it with the bottom-up input, and it uses the error signals to adjust its representation state. In turn, the error signal from this level provides feedback to the convLSTM at the level below it to update that convLSTM. Thus, the convLSTM at each level of representation attempts to explain the bottom-up input while simultaneously trying to align with the top-down expectation from the convLSTM at the next level. The error signal between the prediction and the incoming frame backpropagates through the entire hierarchical network to update all representations, helping it make better predictions with experience. The convLSTM units at each level identify the meaningful spectro-temporal patterns in the local temporal context within a frame as well as the global temporal context of regularities across multiple frames. Figure 1 illustrates the architectural design of PredNet.
We train our PredNet model on an in-house computing cluster with 4-GPU (NVIDIA GeForce GTX 1080 Ti or similar) nodes, using our own PyTorch implementation of PredNet. The model is tested on cross-validation held-out data during training to improve the generalizability of the model.


We conjecture that the PredNet model trained on acoustic music has learned the statistical regularities in the acoustic signals that are experienced by human subjects, and we test whether the network exhibits music perception attributes that reflect the human behavioral data. Individual variations in background, preference, and cultural experience all account for a variety of responses in the music listening experience.  To study musicality perception in a purer form, \cite{Randall2016PrincipalCA} proposed to use a set of randomly-generated pitch sequences that are feature-impoverished but still contain distinct syntactical and organizational features. In their study, 50 monophonic sequences were randomly generated within one key, controlling for timbre, pitch content, pitch range, rhythm, note and sequence length, and loudness. 30 participants listened to these sequences and assigned each clip a musicality rating from one to five, where a rating of "one" represented the "least musical" and "five" represented the "most musical". Ratings showed a clear ranking among the sequences in terms of musicality that is consistent across subjects (Figure 2, top). We use this set of random pitch sequences to evaluate whether human musicality perception is related to the predictability of an auditory sequence based on an experience-dependent model.

\section{Results}
\label{sec:results}

\subsection{Prediction Error is Inversely Proportional to Human Perception of Musicality}

\begin{figure}
    \centering
    \includegraphics[width=1\columnwidth]{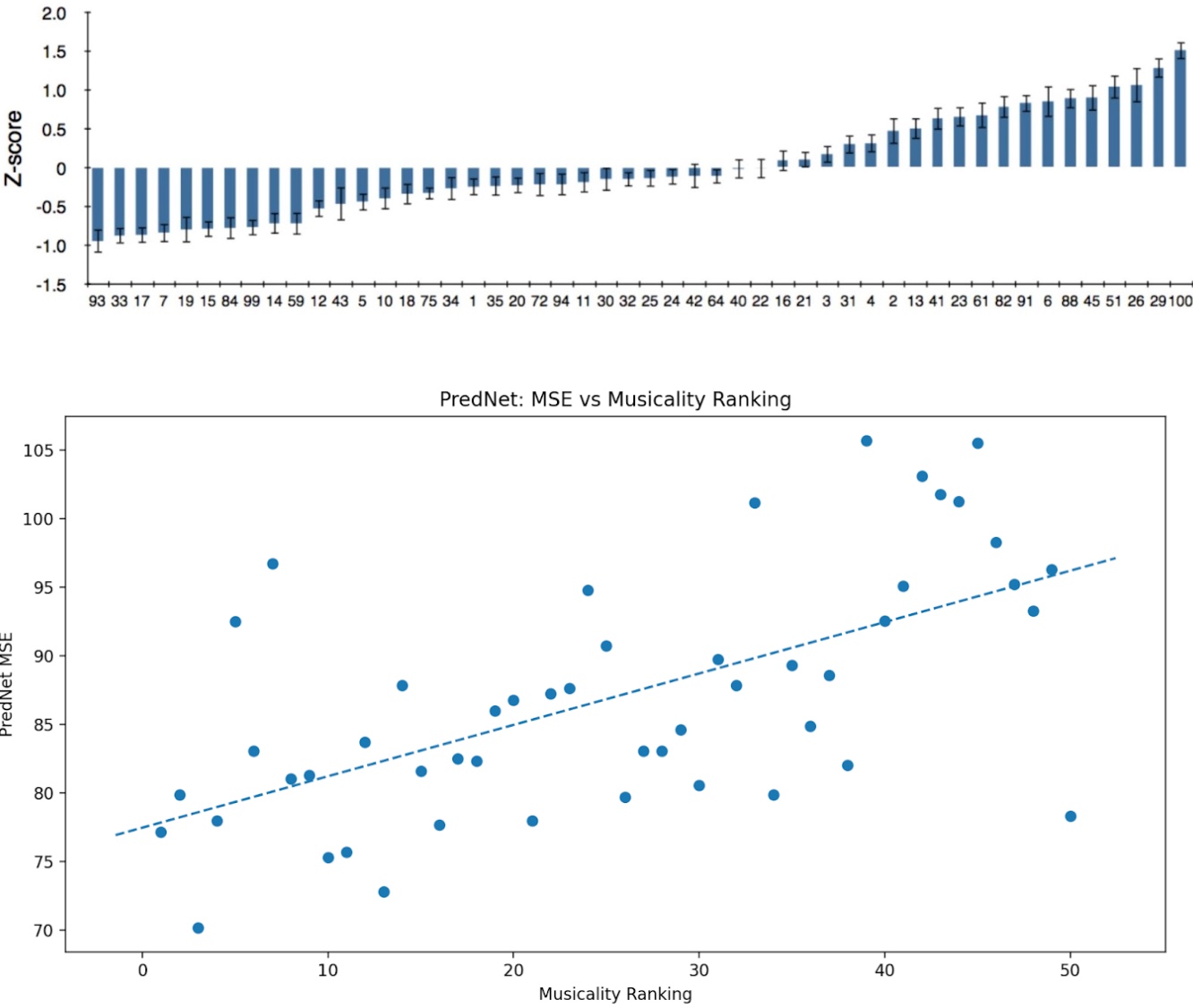}
    \caption[]{Top: Pitch sequences ordered according to the musicality ranking in \cite{Randall2016PrincipalCA}, more musical to the right. Bottom: Averaged prediction errors for the sequences are correlated with the musicality ranking of the sequences.}  
    
\end{figure}

We tested the music-trained PredNet on the 50 randomly-generated pitch sequences in our test dataset \cite{Randall2016PrincipalCA}. We compared PredNet's prediction errors for each sequence, summed over the entire duration of each clip, against the human behavioral data on musicality for that sequence. Figure 2 (bottom) shows that the PredNet's prediction errors indeed are inversely correlated with the musicality ranking $(p < .001, R^2 = .38)$
That is, the more musical sequences evoke lower prediction errors than the more non-musical sequences.

\subsection{Prediction Error Depends on the Amount of Hint or Time Lapse after Pitch Change}

In the test sequences, when one pitch transitions into another, the mel spectrogram first exhibits a transient and then moves to the frequency corresponding to the new pitch. We found the prediction error is inversely proportional to the amount of time lapse after the pitch change. In other words, a longer time  lapse corresponds to a greater "hint" which is available to the network for predicting the next spectrogram frame. The prediction error is the highest when only one time step (11.56 ms) of data is shown after the note change. The prediction error decreases as more time units of the subsequent note become available. At five time units (57.82 ms) of time lapse after note change, the prediction error converges to a minimum and stays at that level afterward until the next pitch change (Figure 3). We compared the average prediction errors of the 10 most musical sequences and that of the 10 most non-musical sequences as a function of time lapse after pitch change, and we found that the prediction errors of the non-musical sequences are consistently stronger than that of the musical sequences (Figure 3). 

\begin{figure}[h]
    \begin{center}
    \includegraphics[width=1\columnwidth, height=140pt]{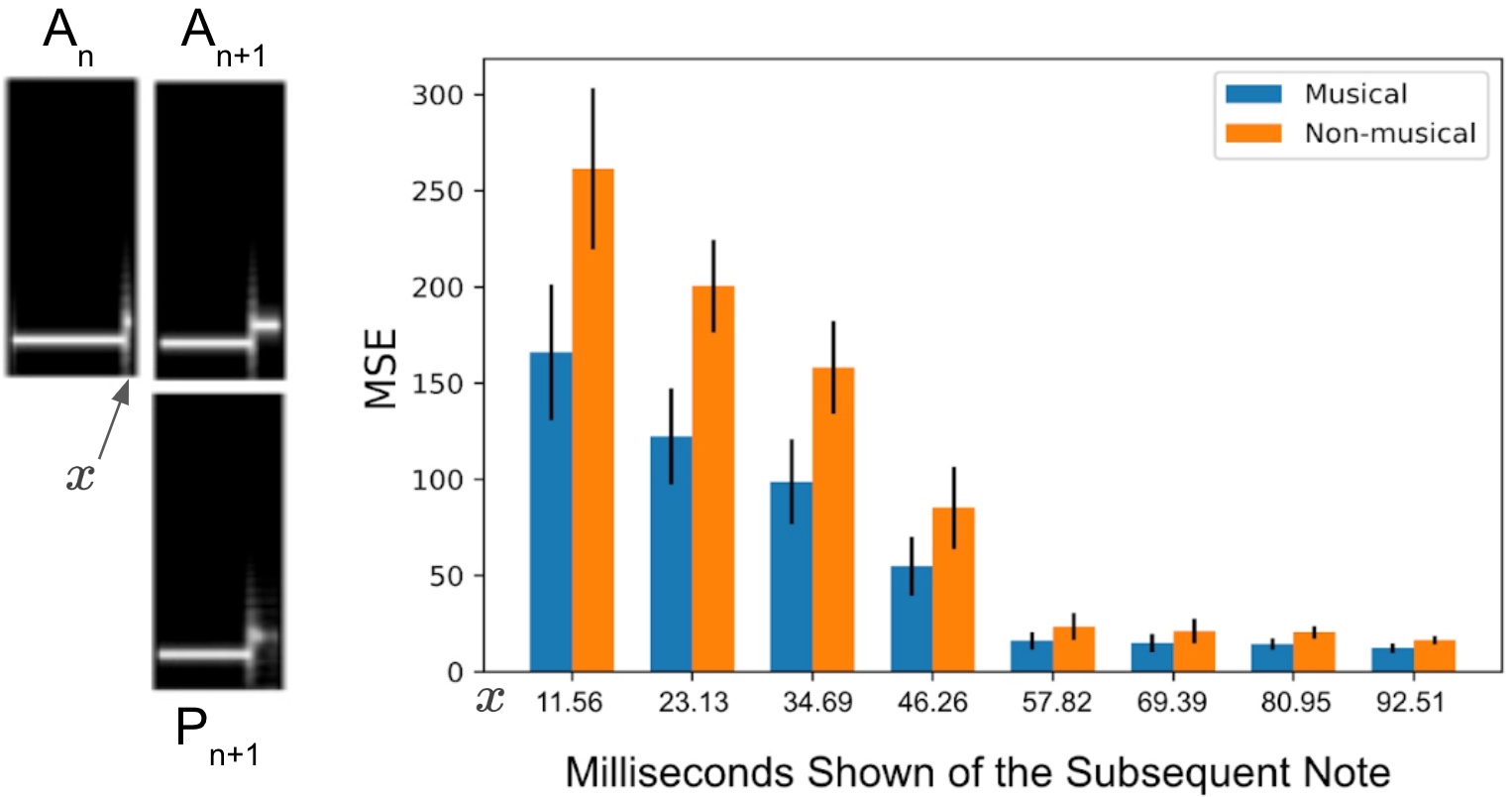}
    \centering

    \end{center}
    \caption{
    Left: Illustration of two input frames ($A_{n}, A_{n+1}$) and the predicted frame ($P_{n+1}$) based on $A_{n}$.
    $x$ is the time lapse after the note transition. 
    Right: 
    Mean squared prediction errors for different values of $x$.  Prediction error decreases with a greater $x$ but is consistently greater for non-musical sequences than for musical sequences up to $x$ = 57.82 ms.}  
    
\end{figure}


\subsection{Prediction Errors Depend on Pitch Interval}

A number of studies have showed that pitch proximity plays an important role in musicality \cite{Randall2016PrincipalCA, huron_2008, Huron_2013, naturebianco}, specifically that musicality is associated with smaller pitch intervals. Larger intervals often trigger discontinuity, which disrupts the listener’s expectation for the progression of a pattern \cite{meyer1973explaining, hadjeres2020vector}. We found a strong correlation ($p < .001$) between the prediction errors at each pitch change to the pitch interval size when the time lapse was one, two, three, and four time units (with 11.56 ms per time unit; examples illustrated in Figure 4). The pitch change size is calculated by the number of frequency bands spanning the interval change in the spectrogram. This observation is consistent with the earlier results in human musicality studies, suggesting the network has learned these statistical regularities in musical signals.



\begin{figure}[h]
    \begin{center}
    \includegraphics[width=1\columnwidth, height=120pt]{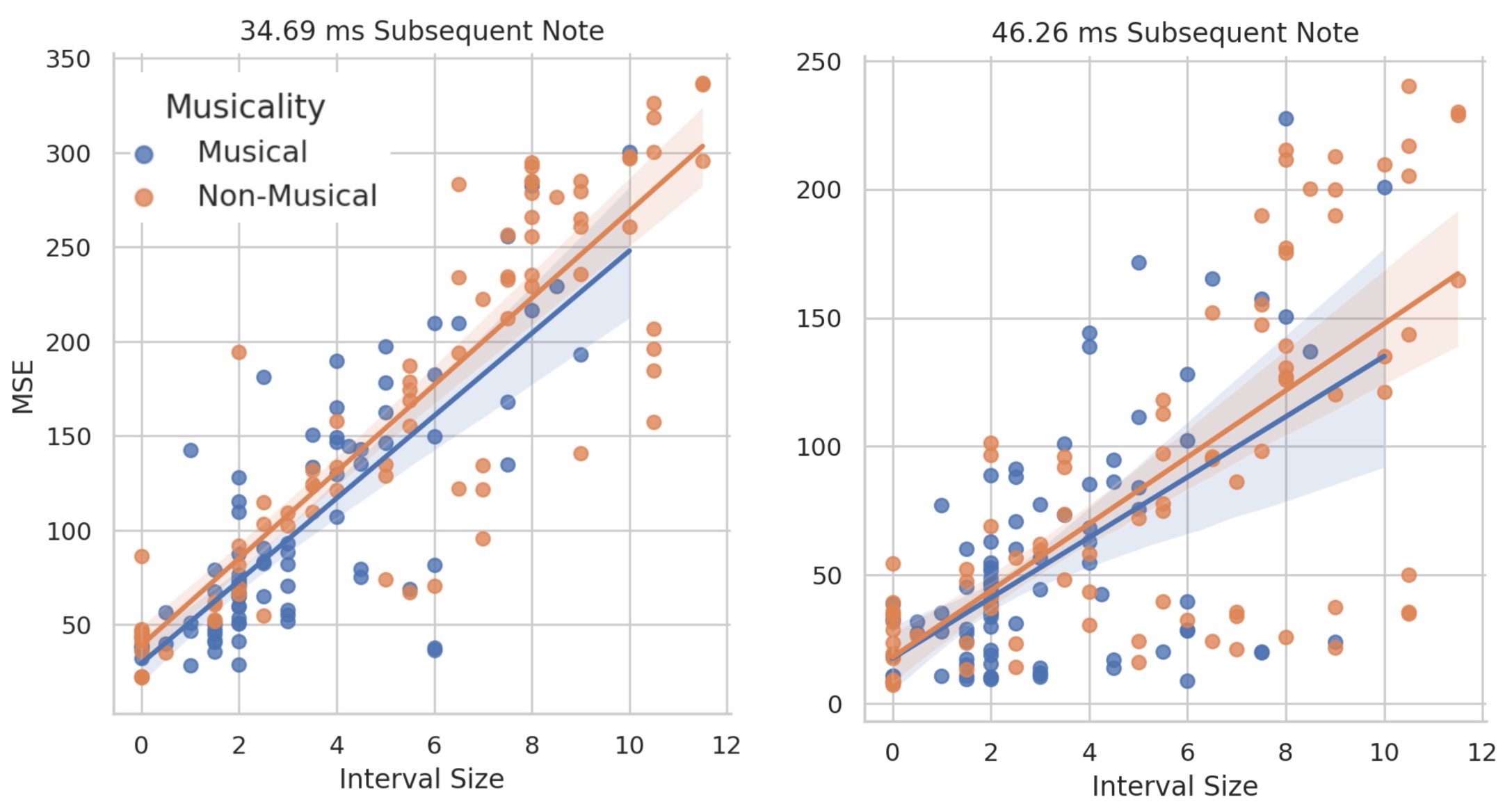}
    \centering
    \end{center}
    \caption{Prediction errors are correlated with pitch interval size for time lapse from $x=11.56$ to $44.26$ ms. Shown are the scatter plots of MSE against interval size at 34.69 ms and 46.26 ms, with $(p < .001, R^2 = .79)$ and $(p < .001, R^2 = .50)$ respectively. } 
\end{figure}

\subsection{Prediction Errors are Sensitive to Temporal Context} 
If the network's prediction is based on a larger temporal context beyond the 508 ms mel spectrogram input frame, then we expect to see an increase in the differential prediction errors between the musical sequences and the non-musical sequences over time (i.e., over the nine note transitions in each sequence). Figure 5 shows an increase in the difference between the normalized prediction errors for the non-musical sequences and the normalized prediction errors for the musical sequences. That is, the later a pitch transition occurs in the sequence, the more predictable the musical sequences are relative to the non-musical sequences. The prediction errors are normalized at each pitch transition by the average pitch change across all the sequences in each group at that pitch change in the sequence. This is designed to compensate for the larger pitch jumps in the non-musical sequences. 





\begin{figure}[htbp]
\centerline{\includegraphics[width=1\columnwidth]{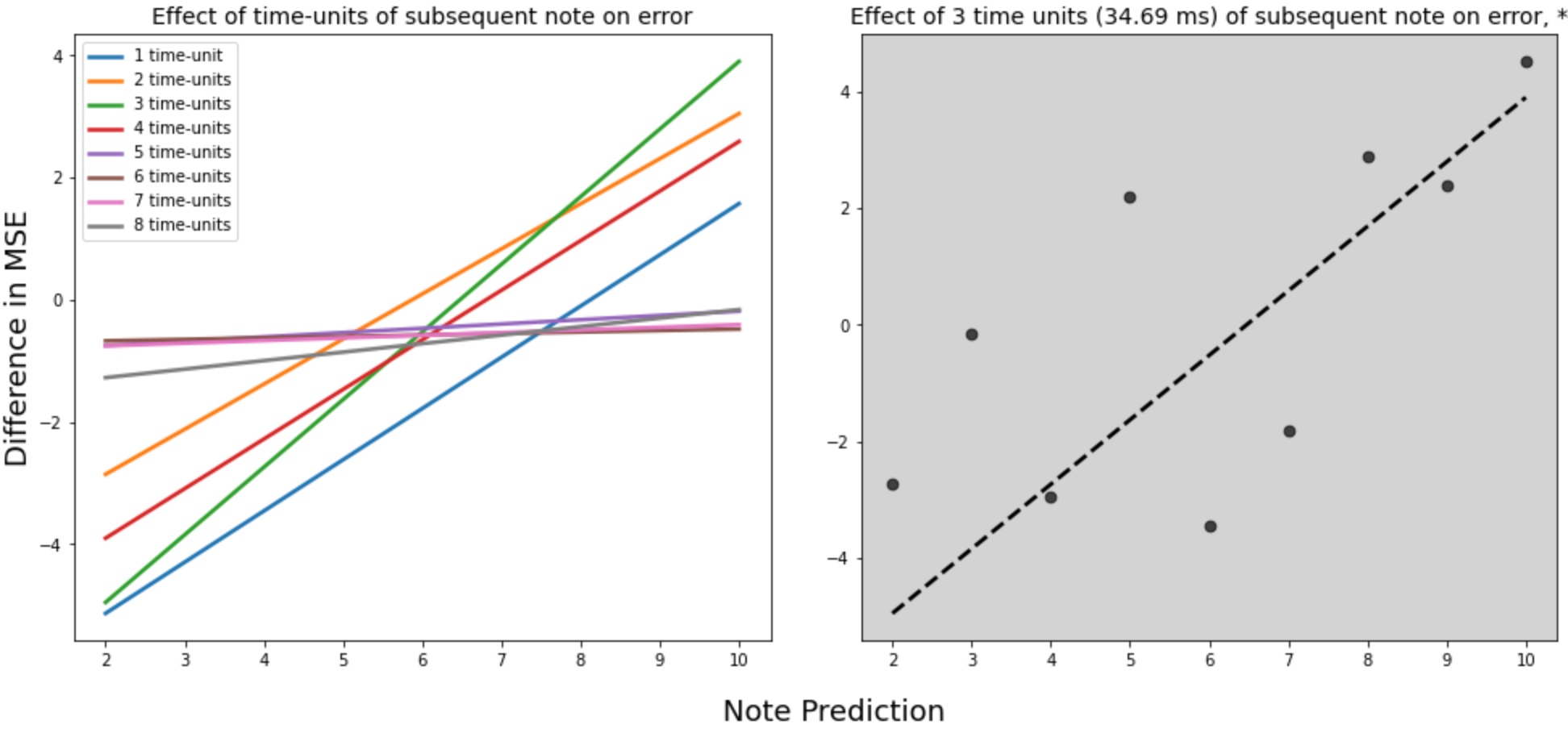}}
\caption{Difference between mean squared prediction error of non-musical sequences and musical sequences increases with time, as the pitch sequence progress, suggesting temporal context improves the prediction of the musical sequences. Left: The regression lines for different values of $x$ (time lapse post-note change). At and after $x$ = 57.82 ms, the difference disappears because sufficient information is available in the mel spectrum to make good predictions for the next frame. Right: The difference of MSE as a function of pitch change location in the sequence at $x$ = 34.69 ms. Dotted line is the regression line $(p = .004, R^2 = .71)$.} 
\label{fig_difference}
\end{figure}

We note that the difference between the non-musical sequences' prediction error and the musical sequences' prediction error over time is significant at time lapses where two, three, and four time units of the subsequent note are shown. The trend exists with 1 time unit of subsequent note but is not statistically significant. This shows the dependence of the prediction errors on a greater temporal context of the pitch sequences. This context effect disappears when greater than four time units of the subsequent note are shown, when there is sufficient hint to converge to the minimum prediction error.

\section{Discussion}


In this study, we showed that a predictive coding neural network trained with musical signals exhibits similar musicality judgment on a set of random pitch sequences to humans. We showed that prediction error is proportional to pitch interval, consistent with earlier studies that have shown the significance of pitch proximity in musicality.
Furthermore, the randomly-generated pitch sequences used are feature-impoverished yet still contain distinct syntactical and organizational features. The sensitivity of the network to temporal context might explain the network's ability to understand the statistical regularities of these structures in the span of multiple seconds. We also found a temporal context effect, which showed that, even when accounting for the difference in average interval size between the non-musical and musical sequences, the model could discriminate between the two categories increasingly well with time. This suggests that the model is sensitive to a more global temporal context. One direction for future research is to characterize the features embedded in the different layers of the neural network to analyze their activity related to syntactic and organizational structures. 

These findings suggest that the PredNet might have indeed learned or captured the statistical regularities of musical signals that our brains have learned, and that characterizing the representational features and motifs in the neural units in this network might help us to uncover the organizational features underlying our music perception.

\vfill\pagebreak

\clearpage

\label{sec:refs}

\bibliographystyle{IEEE}
\bibliography{strings,refs}

\end{document}